\documentclass[aps,prd,onecolumn,floatfix,showpacs,pre,amsmath,amssymb,notitlepage,11pt]{revtex4-2}

\usepackage[dvips]{graphics}
\usepackage{enumitem}

\DeclareMathAlphabet{\mathsfit}{\encodingdefault}{\sfdefault}{m}{sl}
\SetMathAlphabet{\mathsfit}{bold}{\encodingdefault}{\sfdefault}{bx}{sl}

\usepackage[colorlinks,pdfusetitle,urlcolor=blue,citecolor=blue,linkcolor=blue,bookmarksnumbered,plainpages=false]{hyperref}
\usepackage[dvips]{graphics}
\usepackage{geometry}
\DeclareFontEncoding{LS1}{}{}
\DeclareFontSubstitution{LS1}{stix}{m}{n}
\DeclareSymbolFont{symbols2}{LS1}{stixfrak}{m}{n}
\DeclareMathSymbol{\typecolon}{\mathbin}{symbols2}{"25}
 \usepackage[percent]{overpic}

\usepackage{mathptmx}
\usepackage{psfrag}
\usepackage{graphicx}
\usepackage{bm}
\usepackage{amsmath}
\usepackage{trfsigns}
\usepackage{amssymb}
\usepackage{subfigure}
\usepackage[usenames,dvipsnames]{color}
\usepackage{dashrule}
\usepackage{textgreek}
\usepackage[english]{babel}
\usepackage{contour}
\usepackage{placeins}
\usepackage{upgreek,bm}
\usepackage{booktabs}
\usepackage{soul}
\usepackage{color}
\definecolor{dred}{rgb}{.6,.0,0.}
\definecolor{dblue}{rgb}{.0,.0,0.6}
\usepackage{mathtools}

\renewcommand{\vec}[1]{\mathbf{#1}}

\usepackage{xcolor}

\usepackage{lmodern}

\newcommand{\dif}{\mathrm{d}}
\newcommand{\mi}{\textrm{i}} 
\newcommand{\me}{\mathrm{e}}

\begin{document}

\title{Detection of quantum-vacuum field correlations outside the light cone}
\author{Francesca Fabiana Settembrini$^{1}$}
\author{Frieder Lindel$^{2}$}
\author{Alexa Marina Herter$^{1}$}
\author{Stefan Yoshi Buhmann$^{3}$}
\author{Jér\^{o}me Faist$^{1}$}
\affiliation{$^1$ ETH Zurich, Institute of Quantum Electronics, Auguste-Piccard-Hof 1, 8093 Zurich, Switzerland \\ 
$^2$ Physikalisches Institut, Albert-Ludwigs-Universit\"at Freiburg, Hermann-Herder-Stra{\ss}e 3, 79104 Freiburg, Germany\\
$^3$ Institut f\"ur Physik, Universit\"at Kassel, Heinrich-Plett-Stra{\ss}e 40, 34132 Kassel, Germany
}
\maketitle

{\bf 
According to quantum field theory, empty space---the ground state of the theory with all real excitations removed---is not empty at all, but filled with quantum-vacuum fluctuations. Their presence can manifest itself through a series of phenomena such as the Casimir force, spontaneous emission, or dispersion forces. These fluctuating fields possess correlations between space-time points outside the light cone, i.e. between points which are causally disconnected according to special relativity. A counterintuitive consequence is that two initially uncorrelated quantum objects in empty space (the quantum vacuum) which are located in causally disconnected space-time regions, and therefore unable to exchange information, can become correlated.
Using electro-optic sampling, we have experimentally demonstrated the existence of correlations of the vacuum fields for non-causally connected space-time points. This result is obtained by detecting and analyzing vacuum-induced correlations between two 195\,fs laser pulses separated by a time of flight of 470\,fs which propagate through a nonlinear crystal. Our theory reveals the vast majority of the correlations as stemming from space-time points outside the light cone. 
This work marks a first step in analyzing the space-time structure of vacuum correlations in quantum field theory.} \\

\begin{figure}[h]
\centering
\includegraphics[width =0.6\columnwidth]{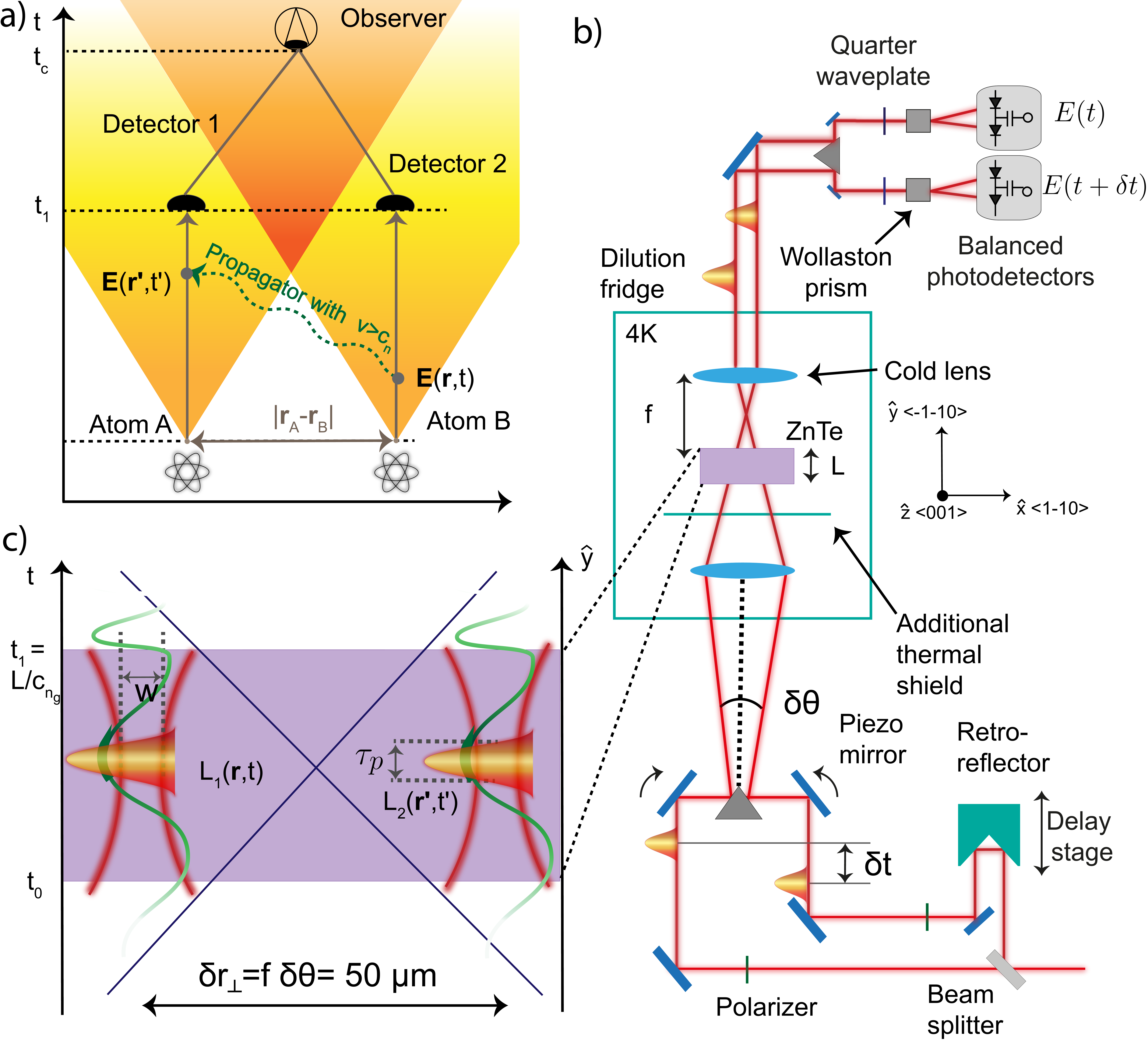}
\caption{ \textit{Nonlocal measurement of quantum-vacuum induced correlations}. (a) Statistical correlations between two atoms A and B can be detected even before an exchange of information between the two can occur. This can be ascribed to their interaction with the quantum vacuum, whose correlation at different space-time points ($\vec{r},t$) and ($\vec{r}^\prime, t^\prime$) is non-vanishing even for non-causally connected regions (in yellow). 
(b) A delay stage and piezo mirrors control the relative spatial $\delta r_{\perp}$ and temporal $\delta t$ distance between the ultrashort probes. The
ZnTe detection crystal is placed in a 4 K environment to isolate it from thermal radiation.~The change in polarization of each probe separately is measured by ellipsometry.
(c) The interaction with the vacuum occurs when the 195fs long pulses enters and leaves the non-linear crystal. } 
\label{fig:conceptual}
\end{figure}

At the core of Einstein's principle of relativity is the fact that two events lying outside the light cone cannot have a causal relationship to each other. In simple words, information can only be carried at the speed of light. Nevertheless, in a very counter-intuitive way, quantum electrodynamics (QED) predicts that two atoms located at a distance $|\vec{r}_A- \vec{r}_B|$ from each other can be correlated in a statistical sense after interacting with the quantum vacuum of the electromagnetic field for a time $\delta  t \leq |\vec{r}_A- \vec{r}_B|/c $, so before they would have time to exchange a photon~\cite{biswas_virtual_1990,valentini_non-local_1991}. Such a surprising result can be interpreted either in terms of swapping of nonlocal correlations from the fluctuating vacuum field to the two atoms or in terms of a nonlocal photon `propagation'~\cite{valentini_non-local_1991}. In this second picture, the Feynman propagator that describes the probability amplitude to create a photon at one location and annihilate it at another has nonzero values outside the forward light cone. \\
The question how vacuum-induced correlations between causally disconnected atoms can be reconciled with special relativity has a long and interesting history. Using the `two-atom' model problem, shown schematically in Fig.~\ref{fig:conceptual} a),  Fermi~\cite{fermi_quantum_1932} concluded already in 1932 that the second atom would react to the state of the first one only after an elapsed time $\delta t \geq |\vec{r}_A- \vec{r}_B|/c$, showing the compatibility between the emerging quantum mechanics and special relativity. Later theoretical work helped firmly establish this result~\cite{hegerfeldt_causality_1994,milonni_photodetection_1995}.
At the technical level, causality is preserved within second-order perturbation theory because the influence of atom A on atom B depends on the commutator of the electric field operator at the two locations of the atoms $\vec{r}_\mathrm{A}$ and $\vec{r}_\mathrm{B}$, respectively, which strictly vanishes outside the light cone in quantum field theory~\cite{peskin2018introduction,schwartz2014quantum}. The correlations on the other hand also involve the anti-commutator of the fields at the two locations. The ground-state expectation value of the latter, similar to the Feynman propagator, is non-vanishing outside the light cone, i.e. 
\begin{align} \label{eq:NonCausCorr}
   \langle \{ \hat{\vec{E}}(\vec{r},t) , \hat{\vec{E}}(\vec{r}^\prime, t^\prime) \} \rangle \neq 0 \quad \mathrm{while} \quad |\vec{r}_\mathrm{A} - \vec{r}_\mathrm{B}| > c|t-t^\prime|.
\end{align}

To summarize, two non-causally connected space-time points cannot influence each other---in keeping with special relativity---but the vacuum can induce correlations between them. Note that as shown in  Fig.~\ref{fig:conceptual} a), a verification of these correlations depends on the comparison of two measurements performed at the two locations which requires a time delay $\delta t \geq |\vec{r}_\mathrm{A} - \vec{r}_\mathrm{B}|/c$ in order to carry the signals to a joint observer.

 These space-time features of the quantum vacuum have recently seen a gain of interest since it was shown that not only classical correlations, but also entanglement could be generated outside the light cone~\cite{reznik_violating_2005,franson_generation_2008}.
 The key idea is to have two probes at two different locations that interact with vacuum field for a finite duration, such that the process occurs in two causally disconnected regions~\cite{reznik_entanglement_nodate}.

Using a technique known as electro-optic sampling, an electric field can be measured inside a non-linear crystal through its interaction with a very short near-infrared laser pulse. 
The unique feature of this technique is that it provides deep subwavelength resolution in space {\em and} time~\cite{Gallot1999} and led to measurements \cite{Riek2015,riek2017subcycle} and proposals~\cite{kizmann2019subcycle,Guedes2019,onoe2021realizing} of vacuum field measurements based on the noise analysis of a single beam.

In an extension of this technique, we have recently shown that electro-optic sampling can be used to measure both the second~\cite{benea2016subcycle} and first order correlations of the electric field at {\em two} different space-time points, and demonstrated that these electric field correlation measurements can be performed even in the limit of the field being in its ground state~\cite{Benea-Chelmus2019}.

We show here that our measurements at two different space-time points also allow to successfully test the existence of field correlations outside the light cone~\cite{valentini_non-local_1991}.
As shown in Fig.~\ref{fig:conceptual}~a) and c), our correlation measurement can be faithfully mapped onto  Fermi’s two-atom system Gedanken experiment. Here, the two atoms that are used to probe the vacuum fluctuations of the electric field are replaced by two laser pulses. Their entering and leaving the non-linear crystal  defines the time during which they will interact with the vacuum fluctuations. After leaving the crystal, they travel without interactions to their respective detectors and allow for the field correlation to be measured.

The working principle of the experimental setup is summarized in Fig.~\ref{fig:conceptual} b). A femtosecond near-infrared pulse generated from a Ti:sapphire laser is split into two identical replicas, which are mutually delayed by a time $\delta t$ and propagate with a relative angle $\delta \theta$. A system of lenses collects and focuses both ultrashort probes into the detection crystal, a 1 mm long $\langle 110\rangle$-cut zinc telluride (ZnTe). The probes polarization is oriented along the $\hat{z}$ axis of the ZnTe crystal and thereby maximizes the electro-optic effect but suppresses all undesired coherent $\chi^{(2)}$ effects \cite{Planken2001}. 
After emerging from the crystal, both ultrashort pulses will have acquired an additional orthogonal electric field component via their interaction with the THz quantum vacuum inside the crystal. A system of ellipsometric balanced detection reads the polarization change of the single probes and retrieves the amplitude of the THz vacuum fluctuations measured inside the nonlinear crystal. From the  electro-optic signal of both beams, the vacuum electric field correlation can be accessed.
The separation between the beams in space is controlled by a symmetric pair of lenses arranged around the ZnTe crystal, such that a relative angle of $\delta\theta$ between the two beams is translated into a separation $\delta r_\perp  = f \delta \theta$ at the focal point of the lens as shown in Fig.~\ref{fig:conceptual} b). To sample non-causally connected regions of the quantum vacuum, the relative distance between the probes has been chosen such that $\delta r_{\perp}=50 \upmu \text{m}>w$, where $w=10\,\upmu$m is their Gaussian beam waist. The THz detection crystal is placed inside a cryostat kept at 4K and shielded to remove all relevant thermal background.

In this way, as we show in Sec. II of the Supplementary Material building upon previous theoretical works \cite{moskalenko2015paraxial,lindel2020theory,lindel2021macroscopic,lindel2021probing}, the experiment measures the following quantity which has been normalized by the detector efficiency such that it has field units:
\begin{equation} 
G_\mathrm{eo}^{(1)}(\delta t, \delta r_\perp)  =  \int \dif t \int_{V_C} \!\!\!\!\! \dif^3 r \int \dif t^\prime \int_{V_C} \!\!\!\!\!  \dif^3  r^\prime  L_1(\vec{r}, t)L_2(\vec{r}^\prime, t^\prime) \langle \{\hat{E}_x(\vec{r}, t) ,\hat{E}_x(\vec{r}^\prime, t^\prime)\} \rangle.
 \label{eq:GMostGeneral}
\end{equation}

 Equation \eqref{eq:GMostGeneral} has an intuitive interpretation which closely connects our experiment to Fermi’s two-atom system in Fig.~\ref{fig:conceptual} a) as discussed before. The two laser pulses can be seen as `vacuum field detectors' moving with the group velocity of the infrared radiation through space and time and thereby `scanning' what we seek to measure over their space-time volumes described by $L_{1,2}$: the vacuum electric field given by the ground-state expectation value of the anti-commutator of the electric field operator $\langle \{ \hat{E}_x(\vec{r}, t) ,\hat{E}_x(\vec{r}^\prime, t^\prime)\} \rangle $. 
 Varying the relative positions of the infrared probes in space and time thus allows to access vacuum correlations between different space-time regions and therefore analyse their space-time structure. \\
 
\begin{figure}[h]
\centering
\includegraphics[width =0.6\columnwidth]{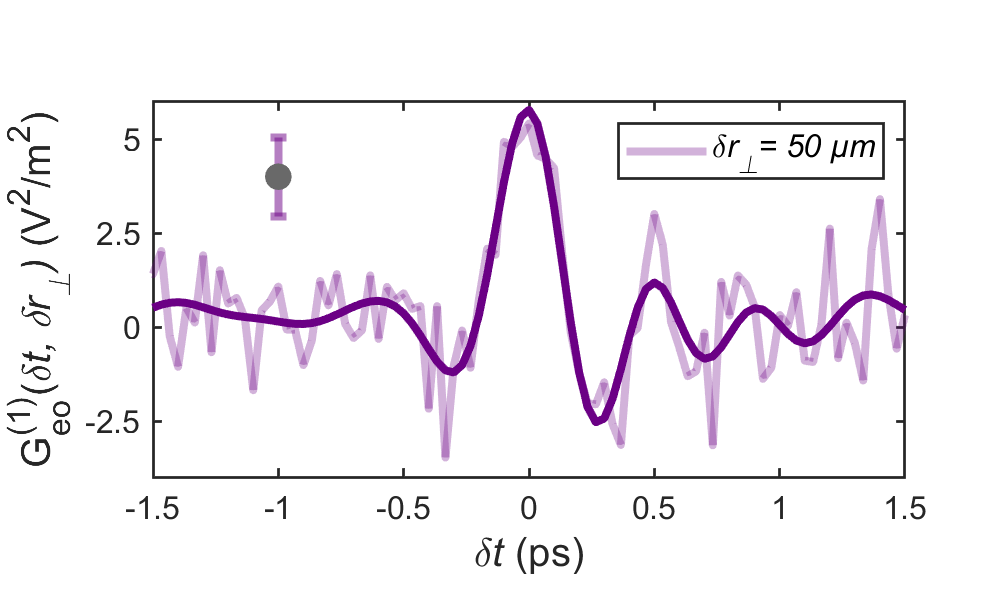}
\caption{ \textit{Experimental measurement}. The electro-optic electric field correlation $G^{(1)}_{eo}(\delta t, \delta r_{\perp})$ on the quantum vacuum is reported as a function of the temporal delay $\delta t$ between the laser pulses. The measurement (faded line) has been performed with an average spacing of $\delta r_{\perp} = 50\, \upmu$m between the probing beams. For better visualization, a 3 THz low-pass filter has been applied to the experimental curve (thick line). The measurement presents a peak-to-peak amplitude of $7.5\,$V$^2/$m$^2$ and a temporal coherence of several picoseconds. The error bar indicates the $2\sigma$ statistical confidence interval.} 
\label{fig:ExperimentalDatacorr}
\end{figure}
 
As shown in Fig.~\ref{fig:ExperimentalDatacorr}, the experimental measurement, performed with a separation of $\delta r_{\perp} = 50\,\upmu$m, yields a very strong signal of $5\,$V$^2$/m$^2$ for zero time delay. The experimental uncertainty is equal to $\sigma = 1.05\,$V$^2$/m$^2$.
In this configuration, the vast majority of the sampled points access non-causal correlations. For instance, the propagation time between the centers of the two probing pulses at $\delta t = 0$ is 470\,fs, which is much longer than the duration $\tau_p = 195\,$fs of the pulse itself. Nevertheless, the two beams still have a finite time duration and transverse width. Moreover, the phase-matching condition does not favor the detection of vacuum fluctuations travelling perpendicularly to the propagation direction of the probing beams (indicated as $\hat{y}$ in Fig.~\ref{fig:conceptual} b)), but rather of those which propagate almost parallel alongside them. \\ 

In order to discriminate the contributions to the electro-optic sampling signal stemming from causally connected space-time regions ($|\vec{r}- \vec{r}^\prime| < |t-t^\prime|c_n $) from those from causally disconnected ones ($|\vec{r}- \vec{r}^\prime| > |t-t^\prime|c_n $), the electric field operator $\hat{E}_x(\vec{r},t )$ is expanded in frequency space to account for the dispersive speed of light inside the crystal. That means we use the experimentally measured refractive index $n(\Omega)$ to obtain the speed of light inside the crystal $c_n = c/n(\Omega)$ for each mode of the vacuum field with frequency $\Omega$. Restricting the space and time integrals in Eq.~\eqref{eq:GMostGeneral} to the causal and non-causal space-time regions, i.e. $|\vec{r}- \vec{r}^\prime| < |t-t^\prime|c/n(\Omega) $ and $|\vec{r}- \vec{r}^\prime| > |t-t^\prime|c/n(\Omega) $, respectively, allows one to split the total electro-optic sampling signal $G^{(1)}_\mathrm{eo}$ into its contribution stemming from causal ($G^{(1)}_\mathrm{eo,c}$) and non-causal vacuum correlations ($G^{(1)}_\mathrm{eo,nc}$) such that $G^{(1)}_\mathrm{eo} = G^{(1)}_\mathrm{eo,nc} + G^{(1)}_\mathrm{eo,c}$. In view of the ultraviolet divergences which are omnipresent in vacuum effects of quantum field theory, this is only meaningful within the frequency bandwidth of the electro-optic detection. See Supplementary Material, Sec. II for details.

\begin{figure}[h]
\centering
\includegraphics[width =0.6\columnwidth]{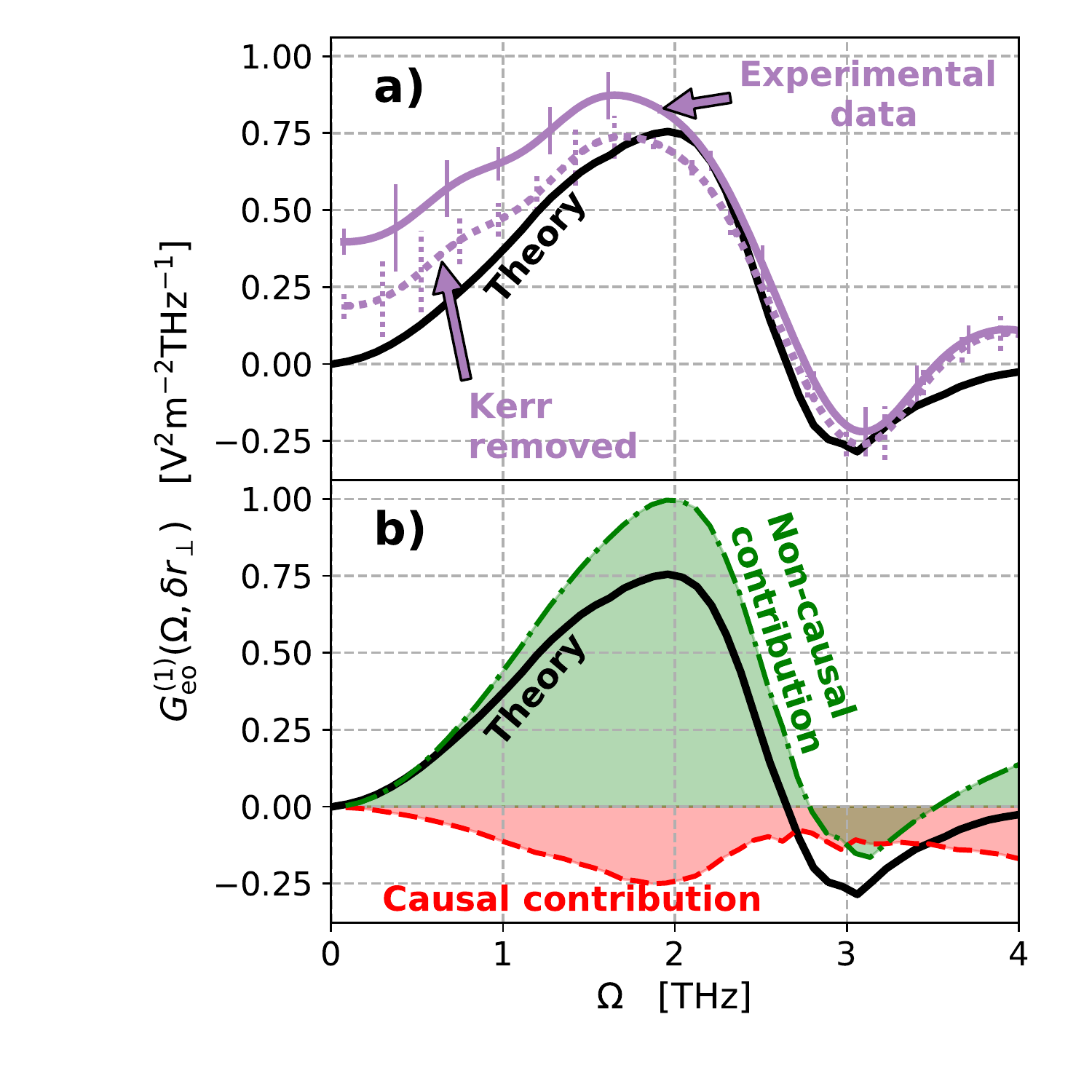}
\caption{
\textit{Causal and noncausal electro-optic signal contributions}. 
(a) The Fourier transform of the apodized measurement raw data shown in Fig.~\ref{fig:ExperimentalDatacorr} (purple) is compared to the theory (black). Additionally, the experimental data with parasitic low-frequency contributions removed is shown by the dotted lines. The experimental error bars are also shown as vertical lines.
(b) The theoretical curve is separated in its causal (red) and non-causal (green) contributions.
}  
\label{fig:ExperimentalData}
\end{figure}

An even more detailed picture that builds upon the benefits of our two-beam technique can be obtained by splitting the total signal into its contributions from different modes of the quantum vacuum and determining causal and non-causal components for each mode. The signal stemming from individual frequency modes is directly obtained, thanks to the Wiener-Khinchin theorem, via a Fourier transform of the correlation along the time delay axis \mbox{$G^{(1)}_\mathrm{eo}(\Omega, \delta r_\perp) = (2\pi)^{-1} \int_{-\infty}^\infty \dif \delta t \, \me^{\mi \delta t \Omega} G^{(1)}_\mathrm{eo}(\delta t, \delta r_\perp)$}. From the theory side, the individual single-mode contributions are derived from Eq.~\eqref{eq:GMostGeneral} by assuming that only a single frequency mode is present in the quantum vacuum such that we define $\hat{E}_{x,\Omega}(\vec{r}, t) = \hat{E}_x(\vec{r}, \Omega) \me^{\mi \Omega t}$ with $\hat{E}_{x}(\vec{r}, t) = \int \dif \Omega \hat{E}_{x,\Omega}(\vec{r}, t)$ to find
 \begin{equation}
     G^{(1)}_\mathrm{eo}(\Omega, \delta r_\perp)  = \int \dif t \int_{V_C} \!\!\!\!\! \dif^3 r \int \dif t^\prime \int_{V_C} \!\!\!\!\!  \dif^3  r^\prime  L_1(\vec{r}, t)L_2(\vec{r}^\prime, t^\prime) \langle \{\hat{E}_{x,\Omega}(\vec{r}, t) ,\hat{E}_{x,\Omega}(\vec{r}^\prime, t^\prime)\} \rangle.
 \end{equation}
As a result, $G^{(1)}_\mathrm{eo}(\Omega, \delta r_\perp)$ allows for a direct comparison between theory and experiment as well as between causal and non-causal contributions in the spectral domain. Its computed value is compared in Fig.~\ref{fig:ExperimentalData}~a) with the same quantity derived from the measured correlation shown in Fig.~\ref{fig:ExperimentalDatacorr} along with the statistical errors of the experimental signal. The raw experimental measurement data has been appropriately filtered in order to reduce noise artefacts (see Supplementary Material, Sec. III). Taking into account also the residual uncertainties in the refractive index, we can consider that theory and experiments agree very well for frequencies above $1$ THz, where the correlation signal peaks. We attribute the additional correlations observed below $ 1$THz to an incomplete removal of parasitic low frequency signals (Supplementary Material, Sec. IV). The experimental data with an estimate of this parasitic contributions removed is shown by the dotted line in Fig.~\ref{fig:ExperimentalData} a).
The causal and non-causal contributions $G^{(1)}_\mathrm{eo,c}(\Omega, \delta r_\perp)$ and $G^{(1)}_\mathrm{eo,nc}(\Omega, \delta r_\perp)$, respectively, are compared in Fig.~\ref{fig:ExperimentalData} b) and clearly show that the signal is dominated by the non-causal contributions. \\ 

In conclusion, we have demonstrated the existence of correlations between fluctuations of the electromagnetic field between non-causally connected space-time regions. If we accept that the state that we measure is the quantum vacuum, a pure state, this implies immediately that the correlations we observe are a proof of entanglement outside the light cone in the quantum vacuum, as predicted theoretically~\cite{reznik_violating_2005}. Our platform provides a first step towards an in-depth analysis of the space-time structure of the quantum vacuum. 

\begin{acknowledgments}
The experimental work was funded by the Swiss National Science Foundation (grant 192330) and the National Centre of Competence
in Research Quantum Science and Technology (QSIT) (grant 51NF40-185902). We also acknowledge the support from the Studienstiftung des deutschen Volkes. We acknowledge the mechanical workshop at ETHZ. We acknowledge the contributions of Dr. Cristina Ileana Benea-Chelmus and Dr. Robert Bennett for their contributions to previous work, Dr. E. Mavrona for the extraction of the refractive index of ZnTe and prof. Renato Renner for insightful comments.  
\end{acknowledgments}

\section*{Authors Contribution}
These authors contributed equally: Francesca Fabiana Settembrini, Frieder Lindel.

J.F., F.F.S. and F.L. conceived the idea for the experiment and its theoretical interpretation. F.F.S and A.H. conducted the measurements. The data analysis was primarily performed by F.F.S and their results were interpreted by F.F.S, F.L and J.F. The theoretical framework was developed by F.L. and S.Y.B. F.L. performed the numerical simulations. J.F. was the scientific supervisor of this work.
The manuscript was written through contributions from all authors. All authors have given approval to the final version of the manuscript.

\bibliography{FinalBib.bib}

\end{document}